\documentclass[%
print,
superscriptaddress,  
amsmath,amssymb,11pt,
]{revtex4}

\usepackage{amsthm}
\usepackage{charter}
\usepackage{graphicx}
\usepackage{dcolumn}
\usepackage{bm}
\usepackage{dsfont}
\usepackage[landscape,papersize={297.1mm,210mm},left=1.9cm,right=1.6cm,top=2.7cm,bottom=2.8cm]{geometry}
\usepackage[                   
            pdfstartview=FitH,
            colorlinks, 
            pdfborder=001,   
            linkcolor=blue,
            anchorcolor=blue,
            citecolor=blue,
            urlcolor=blue
            ]{hyperref}
\newcommand{\ucite}[1]{\textsuperscript{\textsuperscript{\cite{#1}}}}
\begin{document}
\title{Strong superadditivity relations for multiqubit systems}

\author{Xianfei Qi}
\affiliation {School of Mathematics and Statistics, Shangqiu
Normal University, Shangqiu 476000, China}

\author{Ting Gao}
\email{gaoting@hebtu.edu.cn}
\affiliation {School of Mathematical Sciences, Hebei
Normal University, Shijiazhuang 050024, China}

\author{Fengli Yan}
\email{flyan@hebtu.edu.cn}
\affiliation {College of Physics, Hebei
Normal University, Shijiazhuang 050024, China}

\author{Yan Hong}
\affiliation {School of Mathematics and Science, Hebei
GEO University, Shijiazhuang 050031, China}

\begin{abstract}
We investigate the distributions of quantum coherence characterized by superadditivity relations in multipartite quantum systems. General superadditivity inequalities based on the $\alpha$th ($\alpha\geqslant 1$) power of $l_1$ norm of coherence are presented for multiqubit states, which include the existing ones as special cases. Our result is shown to
be tighter than the existing one by a specific example. \\

\end{abstract}


\maketitle

\section{Introduction}
Quantum resource theories \cite{IJMPB2013,PRL115.070503,IC2016,RMP91.025001} provide an extraordinary framework for studying fundamental properties of quantum systems.
Quantum coherence arising from the principle of quantum superposition is an essential feature of quantum mechanics, which marks the departure of quantum world from classical realm. In recent years, the comprehensive formulation of the resource theory of coherence was presented \cite{PRL113.140401,PRL116.120404,PRA95.062327,PRL119.140402,AP2018} (see review papers \cite{RMP89.041003,PR2018}). As an important quantum resource, coherence plays a significant role in many areas such as quantum biology \cite{CP2013}, quantum metrology \cite{NP2011} and thermodynamics \cite{JPA49.143001,CP2016}.

Quantification of coherence is an essential ingredient not only in the theory of coherence but also in the practical application. A rigorous framework for the quantification of coherence is introduced \cite{PRL113.140401} and various computable and meaningful measures of coherence are identified \cite{PRA91.042120,PRA92.022124,PRL115.020403,JPA50.285301,PRL119.150405}. By means of measures of coherence, the issue of the distributions of quantum coherence can be characterized in a quantitative way known as superadditivity relation. For a given bipartite quantum state $\rho_{AB}$, the superadditivity relation is
\begin{equation}\label{1}
\begin{aligned}
C(\rho_{AB})\geqslant C(\rho_{A})+C(\rho_{B}),
\end{aligned}
\end{equation}
where $C$ is a coherence measure, $\rho_{A}=\text{tr}_{B}(\rho_{AB})$ and $\rho_{B}=\text{tr}_{A}(\rho_{AB})$ are the reduced density matrices. In \cite{SR2015}, the superadditivity relation for bipartite quantum states was established based on the relative entropy of coherence. Later, the superadditivity relation was generalized to the case of tripartite pure states \cite{QIP2016}. A sufficient condition to identify the convex roof coherence measures fulfilling the superadditivity relations was provided in \cite{JPA51.414012}. In \cite{QIP2018,IJTP2019}, superadditivity relations in multiqubit systems has been deeply investigated by the use of $l_1$ norm of coherence $C_{l_1}$.

In this paper, we show that superadditivity inequalities related to the $\alpha$th ($\alpha\geqslant 1$) power of $C_{l_1}$ for multiqubit systems can be further improved. We establish a class of tight superadditivity inequalities in multiqubit systems based on $\alpha$th power of $l_{1}$ norm of coherence $C_{l_{1}}$.

\section{Strong superadditivity relations}
Since a resource framework  for quantifying coherence was proposed in \cite{PRL113.140401}, a number of quantities have been proposed to serve as a coherence measure. Among various coherence measures, the $l_{1}$-norm of coherence quantifies coherence in an intuitive way. It can be expressed as
\begin{equation}
\begin{aligned}
C_{l_{1}}(\rho)=\sum\limits_{i\neq j}|\rho_{ij}|,
\end{aligned}
\end{equation}
which is the sum of the absolute value of the off-diagonal entries of the quantum state.

Before we present our main results, we need the following lemmas.

\emph{Lemma 1}. For any $2\otimes 2^{n-1}$ bipartite state $\rho_{AB}$ and $\beta\geqslant 1$, we have
 \begin{equation}
   \begin{aligned}
	C_{l_{1}}^{\beta}(\rho_{AB})\geqslant C_{l_{1}}^{\beta}(\rho_{A})+C_{l_{1}}^{\beta}(\rho_{B}).\label{supadditivity}
   \end{aligned}
 \end{equation}

\emph{Proof}. It follows directly from $C_{l_{1}}(\rho_{AB})\geqslant C_{l_{1}}(\rho_{A})+C_{l_{1}}(\rho_{B})$ for any $2\otimes 2^{n-1}$ bipartite state $\rho_{AB}$ \cite{PRA94.022329}, the monotonicity of the function $f(x)=x^{\beta}$ for $\beta\geqslant 1$, and the inequality $(x+y)^{\beta}\geqslant x^{\beta}+y^{\beta}$ for $x\geqslant 0, y\geqslant 0$ and $\beta\geqslant 1$.

\emph{Lemma 2}\ucite{CTP2019}. Suppose $k$ is a real number satisfying $0<k\leqslant 1$, then for any  $0\leqslant x\leqslant k$, there is
 \begin{equation}
   \begin{aligned}
	(1+x)^{\alpha}\geqslant 1+\frac{(1+k)^{\alpha}-1}{k^{\alpha}}x^{\alpha},
   \end{aligned}
 \end{equation}
for $\alpha\geqslant 1$.

Thus, we have the following theorems.

\emph{Theorem 1}. Suppose a real number $k$ satisfying $0<k\leqslant 1$, any $n$-qubit quantum state $\rho_{A_{1}A_{2}\cdots A_{n}}$ with  $C_{l_{1}}(\rho_{A_{i}})\geqslant \frac{1}{k}C_{l_{1}}(\rho_{A_{i+1}\cdots A_{n}})$ for $i=1,2, \cdots, m$, and $C_{l_{1}}(\rho_{A_{j}})\leqslant \frac{1}{k}C_{l_{1}}(\rho_{A_{j+1}\cdots A_{n}})$ for $j=m+1, \cdots, n-1$, $1\leqslant m\leqslant n-2$ and $n\geqslant 3$, then we have
	\begin{equation}
	\begin{split}
	C_{l_{1}}^{\alpha}(\rho_{A_{1}A_{2}\cdots A_{n}})
&\geqslant C_{l_{1}}^{\alpha}(\rho_{A_{1}})+\left(\frac{(1+k)^{\alpha}-1}{k^{\alpha}}\right)C_{l_{1}}^{\alpha}(\rho_{A_{2}})+\cdots+\left(\frac{(1+k)^{\alpha}-1}{k^{\alpha}}\right)^{m-1}C_{l_{1}}^{\alpha}(\rho_{A_{m}})\\
&\quad +\left(\frac{(1+k)^{\alpha}-1}{k^{\alpha}}\right)^{m+1}[C_{l_{1}}^{\alpha}(\rho_{A_{m+1}})+\cdots+C_{l_{1}}^{\alpha}(\rho_{A_{n-1}})]\\
&\quad +\left(\frac{(1+k)^{\alpha}-1}{k^{\alpha}}\right)^{m}C_{l_{1}}^{\alpha}(\rho_{A_{n}}),
	\end{split}\label{theorem1}
\end{equation}
for all $\alpha\geqslant 1$.

\emph{Proof}. Due to the superadditivity inequality in (\ref{supadditivity}) for $\beta=1$, the monotonicity of the function $f(x)=x^{\alpha}$ for $\alpha\geqslant 1$, and lemma 2, we obtain
\begin{equation}
\begin{split}
C_{l_{1}}^{\alpha}(\rho_{A_{1}A_{2}\cdots A_{n}})
&\geqslant [C_{l_{1}}(\rho_{A_{1}})+C_{l_{1}}(\rho_{A_{2}\cdots A_{n}})]^{\alpha}\\
&= C_{l_{1}}^{\alpha}(\rho_{A_{1}})\left[1+\frac{C_{l_{1}}(\rho_{A_{2}\cdots A_{n}})}{C_{l_{1}}(\rho_{A_{1}})}\right]^{\alpha}\\
&\geqslant C_{l_1}^{\alpha}(\rho_{A_1})\left\{1+\left(\frac{(1+k)^{\alpha}-1}{k^{\alpha}}\right)\left[\frac{C_{l_{1}}(\rho_{A_{2}\cdots A_{n}})}{C_{l_{1}}(\rho_{A_{1}})}\right]^{\alpha}\right\}\\
&= C_{l_1}^{\alpha}(\rho_{A_1})+\left(\frac{(1+k)^{\alpha}-1}{k^{\alpha}}\right)C_{l_{1}}^{\alpha}(\rho_{A_{2}\cdots A_{n}})\\
&\geqslant C_{l_1}^{\alpha}(\rho_{A_1})+\left(\frac{(1+k)^{\alpha}-1}{k^{\alpha}}\right)C_{l_1}^{\alpha}(\rho_{A_2})+\left(\frac{(1+k)^{\alpha}-1}{k^{\alpha}}\right)^{2}C_{l_{1}}^{\alpha}(\rho_{A_{3}\cdots A_{n}})\\
&\geqslant \cdots\\
&\geqslant C_{l_1}^{\alpha}(\rho_{A_1})+\left(\frac{(1+k)^{\alpha}-1}{k^{\alpha}}\right)C_{l_1}^{\alpha}(\rho_{A_2})+\cdots+\left(\frac{(1+k)^{\alpha}-1}{k^{\alpha}}\right)^{m-1}C_{l_{1}}^{\alpha}(\rho_{A_{m}})\\
&\quad+\left(\frac{(1+k)^{\alpha}-1}{k^{\alpha}}\right)^{m}C_{l_{1}}^{\alpha}(\rho_{A_{m+1}\cdots A_{n}}).\label{theorem11}
\end{split}
\end{equation}
Similarly, as $C_{l_{1}}(\rho_{A_{j}})\leqslant \frac{1}{k}C_{l_{1}}(\rho_{A_{j+1}\cdots A_{n}})$ for $j=m+1, \cdots, n-1$, we get

\begin{equation}
\begin{split}
C_{l_{1}}^{\alpha}(\rho_{A_{m+1}\cdots A_{n}})
&\geqslant \left(\frac{(1+k)^{\alpha}-1}{k^{\alpha}}\right)C_{l_{1}}^{\alpha}(\rho_{A_{m+1}})+C_{l_{1}}^{\alpha}(\rho_{A_{m+2}\cdots A_{n}})\\
&\geqslant \left(\frac{(1+k)^{\alpha}-1}{k^{\alpha}}\right)[C_{l_{1}}^{\alpha}(\rho_{A_{m+1}})+\cdots+C_{l_{1}}^{\alpha}(\rho_{A_{n-1}})]+C_{l_{1}}^{\alpha}(\rho_{A_{n}}).\label{theorem12}
\end{split}
\end{equation}

Combining (\ref{theorem11}) and (\ref{theorem12}) gives (\ref{theorem1}), as desired.

 \qed

\emph{Remark 1}. Theorem 4 in \cite{IJTP2019} is the special case $k=1$ of Theorem 1.

\emph{Example}. Let us consider a 3-qubit state
$$|\Psi_{A_{1}A_{2}A_{3}}\rangle=\frac{|0\rangle+|1\rangle}{\sqrt{2}}\otimes |0\rangle\otimes \frac{|0\rangle+3|1\rangle}{\sqrt{10}}.$$
After simple computation, we have $C_{l_1}(\rho_{A_{1}})=1$, $C_{l_1}(\rho_{A_{2}})=0$, $C_{l_1}(\rho_{A_{3}})=3/5$, $C_{l_1}(\rho_{A_{2}A_{3}})=3/5$. Hence, we can choose $k=3/5$. Let $\alpha=2$, then $[(1+k)^{\alpha}-1]/k^{\alpha}=39/9>3=2^{\alpha}-1$. This example shows that our result is better than the one given in \cite{IJTP2019}.

In fact, Theorem 1 can be generalized to the following Theorem.

\emph{Theorem 2}. Let $k$ and $\beta$ be real numbers with $0<k\leqslant 1$ and $\beta\geqslant 1$. For any $n$-qubit quantum state satisfying $C_{l_{1}}^{\beta}(\rho_{A_{i}})\geqslant \frac{1}{k}C_{l_{1}}^{\beta}(\rho_{A_{i+1}\cdots A_{n}})$ for $i=1,2, \cdots, m$, and $C_{l_{1}}^{\beta}(\rho_{A_{j}})\leqslant \frac{1}{k}C_{l_{1}}^{\beta}(\rho_{A_{j+1}\cdots A_{n}})$ for $j=m+1, \cdots, n-1$, $1\leqslant m\leqslant n-2$ and $n\geqslant 3$,  we have
	\begin{equation}
	\begin{split}
	C_{l_{1}}^{\alpha\beta}(\rho_{A_{1}A_{2}\cdots A_{n}})
&\geqslant C_{l_{1}}^{\alpha\beta}(\rho_{A_{1}})+\left(\frac{(1+k)^{\alpha}-1}{k^{\alpha}}\right)C_{l_{1}}^{\alpha\beta}(\rho_{A_{2}})+\cdots+\left(\frac{(1+k)^{\alpha}-1}{k^{\alpha}}\right)^{m-1}C_{l_{1}}^{\alpha\beta}(\rho_{A_{m}})\\
&\quad +\left(\frac{(1+k)^{\alpha}-1}{k^{\alpha}}\right)^{m+1}[C_{l_{1}}^{\alpha\beta}(\rho_{A_{m+1}})+\cdots+C_{l_{1}}^{\alpha\beta}(\rho_{A_{n-1}})]\\
&\quad +\left(\frac{(1+k)^{\alpha}-1}{k^{\alpha}}\right)^{m}C_{l_{1}}^{\alpha\beta}(\rho_{A_{n}}),\label{theorem2}
	\end{split}
\end{equation}
for all $\alpha\geqslant 1$.

\emph{Proof}. Inequality (\ref{theorem2}) can be proved in the same way as (\ref{theorem1}).
\qed

\emph{Remark 2}. When $\beta=1$, Theorem 2 reduces to Theorem 1. Note that, not all coherence measures satisfy superadditivity relation like inequality (\ref{1}) for all quantum states. The method in Theorem 2 can be applied to derive tighter superadditivity inequalities for the case of $x$th($x\geqslant 1$) power of coherence measure satisfying superadditivity relation.

\section{Conclusion}
In this paper, we have focused on the distributions of quantum coherence characterized by superadditivity relations. Tighter superadditivity inequalities related to $\alpha$th ($\alpha\geqslant 1$) power of $l_1$ norm of coherence $C_{l_1}$ for qubit systems are proposed. These new inequalities hold in a tighter way and give rise to finer characterizations of the distributions of coherence. Our result can provide a reference for a better understanding of coherence properties of multipartite quantum systems.
\begin{acknowledgments}
This work was supported by the National Natural Science Foundation of China under Grant Nos: 11947073 and 11701135; the Hebei Natural Science Foundation of China under Grant Nos: A2020205014 and A2018205125, and the Education Department of Hebei Province Natural Science Foundation under Grant No: ZD2020167.
\end{acknowledgments}

\section{Appendix}

\end{document}